\documentclass[10pt,letter,twocolumn]{article}
\usepackage[latin1]{inputenc}
\usepackage{amsmath,amsfonts,amssymb,graphicx}
\usepackage{authblk}

\begin{document}

\title{Adaptive stochastic resonance based on output autocorrelations}
\author[1,2]{Patrick Krauss}
\author[2]{Claus Metzner}
\author[1]{Konstantin Tziridis}
\author[1]{Holger Schulze}
\affil[1]{Experimental Otolaryngology, ENT-Hospital, Head and Neck Surgery, Friedrich-Alexander University Erlangen-Nürnberg (FAU), Germany}
\affil[2]{Department of Physics, Center for Medical Physics and Technology, Biophysics Group, Friedrich-Alexander University Erlangen-Nürnberg (FAU), Germany}
\renewcommand\Authands{ and }

\twocolumn[\begin{@twocolumnfalse} 

\maketitle

\begin{abstract}\textbf{Successful detection of weak signals is a universal challenge for numerous technical and biological systems and crucially limits signal transduction and transmission. Stochastic resonance (SR) \cite{benzi1981mechanism} has been identified to have the potential to tackle this problem, namely to enable non-linear systems to detect small, otherwise sub-threshold signals by means of added non-zero noise. This has been demonstrated within a wide range of systems in physical, technological and biological contexts \cite{wiesenfeld1995stochastic,faisal2008noise}. Based on its ubiquitous importance, numerous theoretical and technical approaches aim at an optimization of signal transduction based on SR \cite{mitaim1998adaptive,mitaim2004adaptive}. Several quantities like mutual information, signal-to-noise-ratio, or the cross-correlation between input stimulus and resulting detector response have been used to determine optimal noise intensities for SR. The fundamental shortcoming with all these measures is that knowledge of the signal to be detected is required to compute them. This dilemma prevents the use of adaptive SR procedures in any application where the signal to be detected is unknown. We here show that the autocorrelation function (AC) of the detector response fundamentally overcomes this drawback. For a simplified model system, the equivalence of the output AC with the measures mentioned above is proven analytically. In addition, we test our approach numerically for a variety of systems comprising different input signals and different types of detectors. The results indicate a strong similarity between mutual information and output AC in terms of the optimal noise intensity for SR. Hence, using the output AC to adaptively vary the amount of added noise in order to maximize information transmission via SR might be a fundamental processing principle in nature, in particular within neural systems which could be implemented in future technical applications.} 
\end{abstract}

\end{@twocolumnfalse}]

\paragraph*{}

Stochastic resonance, a phenomenon first described by Benzi et al. in 1981 \cite{benzi1981mechanism}, has been identified to have the potential to enable non-linear systems to detect arbitrary weak, otherwise sub-threshold signals by means of added noise \cite{collins1996noise,levin1996broadband,gammaitoni1998stochastic}. Already in 1995, Wiesenfeld and Moss have noted in their seminal paper that SR may be demonstrated within a wide range of non-linear systems in physical, technological and biological contexts. In addition, they have demonstrated the existence of an optimal, non-zero intensity for the added noise, allowing maximization of information transmission \cite{wiesenfeld1995stochastic}. To date, the phenomenon receives even increasing attention, especially within the context of experimental and computational neuroscience \cite{faisal2008noise,mino2014effects,douglass1993noise}.
\\
In smart non-linear signal detection systems based on SR, the optimum noise level would have to be continuously adjusted via a feed-back loop, so that the system response in terms of information throughput remains optimal, even if the properties of the input signal change (figure 1). For this processsing principle the term adaptive SR has been coined \cite{mitaim1998adaptive,mitaim2004adaptive,wenning2003activity}. An objective function frequently used in theoretical approaches is the mutual information (MI) or the MI rate, respectively, between the sensor response and the input signal \cite{levin1996broadband,moss2004stochastic,mitaim2004adaptive}. The choice of the MI as objective function is natural since the fundamental purpose of any transducer is to transmit information into a subsequent information processing system. It has been shown previously that the MI as a function of noise intensity has a well-defined peak that indicates the ideal level of noise to be added to the input signal \cite{moss2004stochastic}. However, a fundamental drawback of the MI is the impossibility of calculating it in any application of adaptive SR where the signal to be detected is unknown. The same is true for a number of alternative objective functions, such as the signal-to-noise ratio (SNR) \cite{wiesenfeld1995stochastic,mitaim1998adaptive,levin1996broadband,moss2004stochastic,gammaitoni1998stochastic,douglass1993noise} or the cross-correlation (CC) \cite{collins1995stochastic,collins1996noise} of the input stimulus and the resulting system response. Although finding the optimal level of noise becomes less important with arrays of transducers \cite{collins1995stochastic}, it still remains an unsolved problem for single detector systems.
\\
We argue that this fundamental drawback can be overcome by another objective function, namely the AC of the detector response, which leads to similar or even identical estimates of optimal noise intensities for SR as the aforementioned objective functions, yet with the decisive advantage that no knowledge of the input signal is required (figure 1).
\\
We introduce the concept of the success probability and prove analytically for bipolar signals and a memoryless detector with symmetric thresholds that the output autocorrelation $C_{yy}(\tau\!=\!1)$ yields identical optimum noise levels as the mutual information $I(S;Y)$ and the cross-correlation of signal and output $C_{sy}$ (see methods section). Remarkably, all these funcions can be expressed as strictly monotonous functions of the success probability (for a detailed derivation, see methos section): 
\begin{eqnarray}
C_{sy}&=&2Q - 1 \\
I(S;Y)&=&1 + Q \log_2{Q} + \overline{Q} \log_2{\overline{Q}} \\
C_{yy}(\tau\!=\!1)&=&C_{ss}(\tau\!=\!1) \left( 1 - 4 Q \overline{Q} \right), 
\end{eqnarray}
where $C_{ss}(\tau\!=\!1)$ is the autocorrelation for lag time $\tau\!=\!1$ of the input signal, $Q$ is the success probability, and $\overline{Q}=1-Q$. 
\\
Although these analytical results have been derived in the limited context of a simplified model, we show numerically that the concept holds true for a surprisingly wide range of SR systems. We validated our approach - exemplarily for the comparison of AC and MI - for a variety of models comprising different discrete and continuous input signals, different types of detectors and different signal-to-threshold distances (figures 2 and 3). As aperiodic input signals, three synthetically generated time series were used (figure 2A1-3). A wave file of recorded speech was taken as an example for natural input signals (figure 2A4). A sine signal served as periodic input (figure 2A5). We modeled four types of memoryless detectors with discrete or continuous response functions and symmetric or asymmetric thresholds (figure 3 insets), as well as a leaky integrate-and-fire neuron as a detector with memory. The model shown in figure 2B1 corresponds to the analytically proven model and serves as a validation of the numerical simulation. The cases in figure 2B2-4 are considerably more complex. In all these cases, the MI and the output AC peak at almost identical noise intensities, demonstrating the equivalence of the two measures. Even in the biologically more plausible case of a temporally integrating detector (i.e. the leaky integrate-and-fire neuron \cite{lapique1907recherches,burkitt2006review}) both, the output AC as well as the MI lead to similar optimal noise intensities (figure 2B5). In figure 3 we summarize a large number of investigated models comprising different signal types, different detector types and different signal-to-threshold distances. Here, the optimum noise intensities according to MI are plotted versus the corresponding intensities found by maximizing the output AC. Remarkably both measures are highly correlated (correlation coefficient $r=0.97$).
\\
We conclude that the AC of the detector response may serve as a universal, input independent objective function to estimate the optimal level of noise in SR-like systems. This finding opens up a multitude of possible new technical implementations as it takes adaptive SR from theory to application. In addition, adaptive SR based on output AC may explain a number of natural adaptive processes, especially in neural systems where this mechanism would be particularly plausible as the computation of the AC may easily be implemented in neural networks \cite{licklider1951duplex}. Finally, malfunctioning adaptive SR in neural sensory systems may be responsible for pathologic conditions like neuropathic pain or tinnitus, where the unsuccessful attempt to compensate for peripheral receptor damage leads to maladaptive central neural plasticity.

\section*{Methods}
{\small

\paragraph*{Mutual Information}

The mutual information $I(S;Y)$, a measure frequently used in probability theory and information theory, quantifies the mutual dependence of two random variables $S$ and $Y$. It determines how similar the joint distribution $p(s,y)$ is compared to the product of the factored marginal distributions $p(s)p(y)$ and is derived from Shannon's entropy \cite{shannon1959mathematical}.
\begin{equation}
  I(S;Y)= \sum_{s}\sum_{y}\; p(s,y)\;\log_2\left( \frac{ p(s,y)}{p(s)p(y)} \right),
\label{MI}
\end{equation}
where $p(s,y)$ is the joint probability distribution function of $S$ and $Y$, and $p(s)$ and $p(y)$ are the marginal probability distribution functions of $S$ and $Y$ respectively. For continuous random variables, the summation is replaced by a double integral:
\begin{equation}
  I(S;Y) = \int_S \int_Y p(s,y) \log_2{ \left(\frac{p(s,y)}{p(s)\,p(y)} \right) } \; dy \,ds, 
\label{MI2}
\end{equation}
where $p(s,y)$ is now the joint probability density function of $S$ and $Y$, and $p(s)$ and $p(y)$ are the marginal probability density functions. The natural unit of $I(S;Y)$ is $bits$, however in some cases it might be more convenient to divide the total mutual information by the time or by the number of spikes within the observed spike train and thus derive mutual information rates $R(S;Y)$ measured in $bits \, s^{-1}$ or $bits \, spike^{-1}$. The choice of the mutual information as an objective function is natural, because the fundamental purpose of any sensor is to transmit information into a subsequent information processing system. Indeed, it has been shown by several authors \cite{levin1996broadband,moss2004stochastic,mitaim2004adaptive} that, within the context of stochastic resonance, $I(S;Y)$ as a function of the variance $\sigma^2$ of the added noise has a maximum that indicates the optimal level of noise.

\paragraph*{Signal-to-noise ratio}

The signal-to-noise ratio (SNR) in dB is given by the standard formula \cite{zhou1990escape,wiesenfeld1995stochastic,douglass1993noise,levin1996broadband}
\begin{equation}
\mbox{SNR}_{dB}=10\,\log_{10}{(S/N(f))}
\label{SNR}
\end{equation}
where $N(f)$ is the amplitude of the noise power density at the stimulus frequency when presented alone, and $S$ (signal power) is the area under the signal peak above the noise in the joint presentation case \cite{levin1996broadband}. Remarkably, the above mentioned mutual information rate $R$ is directly related to the non-logarithmic SNR through the formula
\begin{equation}
R=\int_0^{\infty} \log_2{(1+\frac{S}{N(f)})} df
\label{SNR2MI}
\end{equation}
for gaussian distributions of signal and noise \cite{shannon1959mathematical}.

\paragraph*{Cross-correlation}

The normalized cross-correlation for time lag $\tau=0$ of stimulus $S$ and detector response $Y$ is defined as 
\begin{equation}
C_{sy} = \frac{\left\langle 
(s_{t\!}-\overline{s})\;(y_t-\overline{y})
\right\rangle_t}
{\sqrt{\left\langle (s_t-\overline{s})^2 \right\rangle_t \left\langle (y_t-\overline{y})^2 \right\rangle_t}},
\label{CC}
\end{equation}
where $\overline{s}$ and $\overline{y}$ are the means and $\left\langle \cdot \right\rangle_t$ indicates averaging over time.

\paragraph*{Output autocorrelation}

The standard autocorrelation as a function of the time lag $\tau$ is defined as
\begin{equation}
C_{yy}(\tau) = \frac{\left\langle 
(y_{t\!+\!\tau}-\overline{y})\;(y_t-\overline{y})
\right\rangle_t}
{\left\langle (y_t-\overline{y})^2 \right\rangle_t},
\label{AC}
\end{equation}
where $\overline{y}$ is the mean and $\left\langle \cdot \right\rangle_t$ indicates averaging over time. We note that for most applications (and discrete time steps) it is sufficient to consider only one time step, i.e. $\tau=1$. However, for more complex signals, e.g. streams of n-bit words, it might be beneficial to calculate $C_{yy}(\tau)$ for a number of different subsequent lag times. In order to derive a single value from the function, the root mean square (rms) of the autocorrelation function
\begin{equation}
\mbox{RMS}(C_{yy})=\sqrt{\frac{1}{N_{\tau}}\sum_{\tau}(C_{yy}(\tau))^2}
\label{ACRMS}
\end{equation}
may be calculated, where $N_{\tau}$ is the total number of different time lages.

\paragraph*{Success probability}

The output of a memory-less sensor can be described by a conditional probability distribution $p(y_t|s_t,n_t)$, which includes deterministic behaviour as a special case. Assuming statistically independent noise with distribution $p(n_t)$, the signal transmission properties of the sensor are given by $p(y_t|s_t)=\sum_{n_t}p(y_t|s_t,n_t)p(n_t)$. Ideally, the sensor output should be equal to the input signal, $y_t\!=\!s_t$, so that $p(y_t|s_t)=\delta_{y_t,s_t}$. It is therefore meaningful to quantify the performance of a sensor by the success probability
\begin{equation}
Q=p(y_t\!=\!s_t),
\end{equation}
which is expected to peak at some optimum noise level within the context of stochastic resonance.

\paragraph*{Analytical model}

In general, the momentary response of a SR-sensor can depend on the history of internal states of the system, as is the case in integrate-and-fire-neurons. For simplicity, in the analytical model we only consider memory-less sensors, which respond to the present input signal $s_t$ and noise value $n_t$ independently from their former activity states.

We consider a bipolar stochastic sensor in which both the input signal $s_t$ and the sensor output $y_t$ can only take on the values $-1$ and $+1$. The noise values $n_t$, however, are continuous gaussian random numbers with variance $\sigma^2$ and mean $\mu\!=\!0$. We further assume that these two values appear in the input signal with equal probalility, $p(s_t\!=\!-1)\!=p(s_t\!=\!+1)\!=\!0.5$. By assuming two symmetric detection thresholds $\pm\theta$ (figure 3 lower right inset), together with symmetric white noise, it can be assured that the distribution of sensor outputs $p(y_t\!=\!-1)\!=p(y_t\!=\!+1)\!=\!0.5$ is also symmetric, so that the mean, variance and entropy of $y_t$ remain constant even if the noise level is changed. Hence, the expressions for $I(S;Y)$ and $C_{yy}(\tau)$ can be slightly simplified. In particular, the autocorrelation can be reduced to the non-normalized form $C_{yy}(\tau)\propto \left\langle y_t y_{t+\tau} \right\rangle$, and, furthermore, will be considered only for lagtime $\tau\!=\!1$.

The sensor adds the noise $n_t$ to the binary input signal $s_t$. If $s_t\!+\!n_t$ exceeds the upper threshold $\theta$, the output $y_t$ is $+1$, if $s_t\!+\!n_t$ falls below the lower threshold $-\theta$, output $y_t$ is $-1$. For $s_t\!+\!n_t\!\in\left[-\theta,+\theta \right]$, the output is chosen randomly between the two binary values $+1$ and $-1$. 

We are interested in the case of a threshold $\theta\!>\!1$ which exceeds the signal amplitude, so that without the assistance of added noise the signal cannot be detected. Adding a random noise value $n_t$ to a (say) positive input signal $s_t$ can have three possible effects. If we consider the noise to be sufficiently positive to lift the signal beyond the upper threshold, then the success probability $Q=p(y_t\!=\!+1 \;|\; s_t\!=\!+1)=p(y_t\!=\!-1 \;|\; s_t\!=\!-1)$ will be increased. Alternatively, if the noise happens to be strongly negative and draws the positive signal below the lower threshold $-\theta$ then the success probability $Q$ will be decreased. The third possibility is that $s_t\!+\!n_t$ remains sub-threshold. Such cases make the signal transmission neither better nor worse. 

It is intuitively clear that small noise levels will increase $Q$, but as soon as a considerable fraction of momentary noise levels $n_t$ exceeds $2\!+(\theta\!-\!1)$, the success probability $Q$ will fall again. In our case it is given by $Q = Q(\sigma) = \frac{1}{2}+\frac{1}{2}\left[ \;W(\frac{\theta\!+\!1}{\sigma}) - W(\frac{\theta\!-\!1}{\sigma})\;\right],$
where $W(x)\!=\!\frac{1}{2} \mbox{erf}(\frac{x}{\sqrt{2}})$ is a slightly rescaled error function (see \emph{Derivation of success probability} for a detailed derivation). As a function of the noise level $\sigma$, the success probability has a well-defined maximum.

In this sensor model, the mutual information $I(Y;S)$ can be expressed as a strictly increasing function of the success probability: $I(Q) = 1 + Q \log_2 Q + (1\!-\!Q) \log_2(1\!-\!Q)\;$ (see \emph{Derivation of mutual information} for a detailed derivation).

Since both $I$ and $Q$ require access to the sub-threshold signal $s_t$, we turn to the autocorrelation function $C_{yy}$ of the sensor output. Since the mean $\overline{y}$ of $y_t$ is zero and its variance constant, we can use a non-normalized version of equation(6). Furthermore, we restrict our analytical consideration to a single lag-time $\tau\!=\!1$, defining $\mbox{C}=\left\langle y_t y_{t+1} \right\rangle$. The modulus of this quantity, too, can be expressed as a strictly increasing function of the success probability: $|C(Q)| = |\left\langle s_t s_{t+1} \right\rangle|\;\left[ 1-4Q(1\!-\!Q)\right]$, where $\left\langle s_t s_{t+1} \right\rangle$ are the input correlations (see \emph{Derivation of output autocorrelation} for a detailed derivation).

\paragraph*{Derivation of success probability}

\noindent
The normalized Gaussian distribution with zero-mean and standard deviation $\sigma$ is given by

\begin{equation}
g(x,\sigma)=\frac{1}{\sqrt{2\pi}\sigma}\; e^{-\frac{1}{2}(x/\sigma)^2}
\end{equation}

\noindent
For later convenience, we define a function $W(x)$ via

\begin{equation}
W\left(\frac{z}{\sigma}\right) = \int_0^{z} \!\!g(x,\sigma) \;dx = \frac{1}{2}\;\mbox{erf}\left(\frac{1}{\sqrt{2}}\;\frac{z}{\sigma}\right),
\end{equation}

\noindent
where $\mbox{erf}(x)=\frac{2}{\sqrt{\pi}} \int_0^{x} \!\! e^{-t^2}dt$ is the error function.

\vspace{0.5cm}
\noindent
The success probability $Q$ is given by

\begin{eqnarray}
Q &=& p(y_t\!=\!+1 | s_t\!=\!+1)=\nonumber\\
&=& \frac{1}{2}\cdot p(\;-\theta-1 < n_t < \theta -1\;)+\nonumber\\
&+& p(\;n_t>\theta-1\;)
\end{eqnarray}

\noindent
The factor $\frac{1}{2}$ accounts for the stochastic output of the unit in the case when $s_t\!+\!n_t$ is sub-threshold. We can now express the probabilities as integrals over Gaussians:

\begin{eqnarray}
Q &=& \frac{1}{2}\cdot\left( \int_0^{\theta-1}\!\!g(x,\sigma)dx + \int_0^{\theta+1}\!\!g(x,\sigma)dx \right) + \nonumber\\
&+& \left( \frac{1}{2} - \int_0^{\theta-1}\!\!g(x,\sigma)dx \right)
\end{eqnarray}

\noindent
Next we use the function $W(x)$ defined above:

\begin{eqnarray}
Q &=& \frac{1}{2}\cdot\left( W\left(\frac{\theta-1}{\sigma}\right) + W\left(\frac{\theta+1}{\sigma}\right) \right) + \nonumber\\
&+& \left( \frac{1}{2} - W\left(\frac{\theta-1}{\sigma}\right) \right)=\nonumber\\
&=& \frac{1}{2}\;+\;\left[\; W\left(\frac{\theta+1}{\sigma}\right) - W\left(\frac{\theta-1}{\sigma}\right)\; \right]
\end{eqnarray}

\paragraph*{Derivation of mutual information}

\noindent
The mutual information of the detector output and the input signal is defined as

\begin{eqnarray}
I(Y;S)&=& \sum_{y,s}\; p(y,s)\;\log_2\left( \frac{ p(y,s)}{p(y)p(s)} \right)=\nonumber\\
&=&  \sum_{y,s}\; p(y|s)p(s)\;\log_2\left( \frac{p(y|s)p(s)}{p(y)p(s)} \right)=\nonumber\\
&=&  \sum_{y,s}\; p(y|s)(1/2)\;\log_2\left( \frac{p(y|s)(1/2)}{(1/2)(1/2)} \right)=\nonumber\\
&=&  \frac{1}{2}\;\sum_{y,s}\; p(y|s)\;\log_2\left( 2 p(y|s)\right).
\end{eqnarray}

\noindent
We explicitly go through all four terms:

\begin{eqnarray}
2I(Y;S)&=&  \sum_{y,s}\; p(y|s)\;\log_2\left( 2 p(y|s)\right)=\nonumber\\
&=& p(y\!=\!-1|s\!=\!-1) \;\log_2\left( 2 p(y\!=\!-1|s\!=\!-1)\right)+\nonumber\\
&+& p(y\!=\!-1|s\!=\!+1) \;\log_2\left( 2 p(y\!=\!-1|s\!=\!+1)\right)+\nonumber\\
&+& p(y\!=\!+1|s\!=\!-1) \;\log_2\left( 2 p(y\!=\!+1|s\!=\!-1)\right)+\nonumber\\
&+& p(y\!=\!+1|s\!=\!+1) \;\log_2\left( 2 p(y\!=\!+1|s\!=\!+1)\right)=\nonumber\\
&=& Q \;\log_2\left( 2 Q\right)+\nonumber\\
&+& (1-Q) \;\log_2\left( 2 (1-Q)\right)+\nonumber\\
&+&(1-Q) \;\log_2\left( 2 (1-Q)\right)+\nonumber\\
&+& Q \;\log_2\left( 2 Q\right).
\end{eqnarray}

\noindent
Therefore

\begin{eqnarray}
I(Y;S)&=&  Q \;\log_2\left( 2 Q\right)+(1-Q) \;\log_2\left( 2 (1-Q)\right)\nonumber\\
&=& 1+Q \;\log_2(Q) + (1-Q) \;\log_2(1-Q).
\end{eqnarray}

\paragraph*{Derivation of output autocorrelations in the analytical model}

\noindent
The temporal correlations of the input signal can be expressed by the probability $q=p(s_1\!=\!+1,s_0\!=\!+1)$ in the following way: 

\begin{eqnarray}
&\;& \left\langle s_{t+1}\; s_t\right\rangle = \left\langle s_1\; s_0\right\rangle =\nonumber\\
&=&  \sum_{s_0,s_1}\; p(s_1,s_0)\; (s_1\;s_0)=\nonumber\\
&=& p(s_1\!=\!-1|s_0\!=\!-1)p(s_0\!=\!-1) \left[(-1)(-1) \right] +\nonumber\\
&+& p(s_1\!=\!-1|s_0\!=\!+1)p(s_0\!=\!+1) \left[(-1)(+1) \right] +\nonumber\\
&+& p(s_1\!=\!+1|s_0\!=\!-1)p(s_0\!=\!-1) \left[(+1)(-1) \right] +\nonumber\\
&+& p(s_1\!=\!+1|s_0\!=\!+1)p(s_0\!=\!+1) \left[(+1)(+1) \right] =\nonumber\\
&=& q\;(1/2)\left[1 \right] +\nonumber\\
&+& (1-q)\;(1/2) \left[-1\right] +\nonumber\\
&+& (1-q)\;(1/2) \left[-1\right] +\nonumber\\
&+& q\;(1/2) \left[1 \right] = 2q-1.
\end{eqnarray}

\noindent
The temporal correlations in the output signal are given by

\begin{eqnarray}
C_{yy}(\tau=1)&=& \left\langle y_{t+1}\; y_t\right\rangle = \left\langle y_1\; y_0\right\rangle =\nonumber\\
&=&  \sum_{y_0,y_1}\; p(y_1,y_0)\; (y_1\;y_0).
\end{eqnarray}

\noindent
Consider for example the probability $p(y_1=+1,y_0=+1)$. There are four different chains of events which can produce a sequence of two successive $+1$'s in the output signal:

\begin{eqnarray}
&\;&p(y_1\!=\!+1,y_0\!=\!+1) =\nonumber\\
&=&p(y_1\!=\!+1|s_1\!=\!-1) p(s_1\!=\!-1|s_0\!=\!-1) \cdot \nonumber\\
&\cdot& p(y_0\!=\!+1|s_0\!=\!-1) p(s_0\!=\!-1)+\nonumber\\
&+&p(y_1\!=\!+1|s_1\!=\!-1) p(s_1\!=\!-1|s_0\!=\!+1) \cdot \nonumber\\
&\cdot& p(y_0\!=\!+1|s_0\!=\!+1) p(s_0\!=\!+1)+\nonumber\\
&+&p(y_1\!=\!+1|s_1\!=\!+1) p(s_1\!=\!+1|s_0\!=\!-1) \cdot \nonumber\\
&\cdot& p(y_0\!=\!+1|s_0\!=\!-1) p(s_0\!=\!-1)+\nonumber\\
&+&p(y_1\!=\!+1|s_1\!=\!+1) p(s_1\!=\!+1|s_0\!=\!+1) \cdot \nonumber\\
&\cdot& p(y_0\!=\!+1|s_0\!=\!+1) p(s_0\!=\!+1)=\nonumber\\
&=&  (1-Q)\;q\; (1-Q)\; (1/2)+ \nonumber\\
&+&  (1-Q)\;(1-q)\; Q\; (1/2)+ \nonumber\\
&+&  Q\;(1-q)\; (1-Q)\; (1/2)+ \nonumber\\
&+&  Q\;q\; Q\; (1/2)= \nonumber\\
&=& \frac{q}{2} + (2q-1) Q (1-Q) =: A.
\end{eqnarray}

\noindent
For symmetry reasons, $p(y_1=-1,y_0=-1)=p(y_1=+1,y_0=+1) = A$. In the same way, $p(y_1=+1,y_0=-1)=p(y_1=-1,y_0=+1) = B$.

\vspace{0.5cm}
\noindent
Since $\sum_{y_0,y_1}\! p(y_1,y_0)=1=2A+2B$, it follows that $B=\frac{1}{2}-A$.

\vspace{0.5cm}
\noindent
Knowing all four joint probabilities, we can proceed to compute the temporal correlations in the output signal:

\begin{eqnarray}
C_{yy}(\tau=1)&=& \left\langle y_1\; y_0\right\rangle =\nonumber\\
&=& A(-1)(-1) + B(-1)(+1) + \nonumber\\
&+& B(+1)(-1) + A(+1)(+1)=\nonumber\\
&=& 2A-2B=\nonumber\\
&=& (2q-1)\left[\;1-4Q(1-Q)\;\right]=\nonumber\\
&=& \left\langle s_{t+1}\; s_t\right\rangle \;\left[\;1-4Q(1-Q)\;\right].
\end{eqnarray}

\paragraph*{Input signals for numerical simulations}

Both, synthetically generated as well as natural signals were used in numerical simulations. As a discrete input signal, a correlated, bipolar string $s_t\in\{-1\!,\!+1\}$ was generated, in which the probability of successive values being identical was $\mbox{Prob}(s_t\!=\!s_{t-1})\!=\!0.7$. As continuous signals in order of increasing complexity we used: first, a sine waveform signal with constant frequency and amplitude; second, an aperiodic time series derived from the variable $x(t)$ of the Roessler attractor [16]
\begin{eqnarray}
\dot{x}&=&-(y+z)\nonumber\\
\dot{y}&=&x + ay\nonumber\\
\dot{z}&=&b + (x-c)z,\nonumber\\
\end{eqnarray}
with parameters $a=0.15$, $b=0.2$ and $c=7.1$; third, the aperiodic random Ornstein-Uhlenbeck process [17] $\dot{x}=-\frac{1}{\tau}x+\epsilon \;\xi(t)$, where $\xi$ is an independent normally distributed random variable, $\tau$ is the correlation time and $\epsilon$ is the noise amplitude; fourth, wave files of speech, music and natural sounds have been recorded. All synthetically generated signals were computed by numerically integrating the differential equations using fourth order Runge Kutta method.

\paragraph*{Sensor models for numerical simulations}

Four different memory-less sensor models were implemented by combining symmetric and asymmetric thresholds, with discrete and continuous sensor output functions. In the symmetric models, there exist two thresholds $+\theta$ and $-\theta$. Without added noise, the sensor output is zero for $|s_t|<\theta$, $s_t\!-\!\theta$ for $s_t>\theta$, and $s_t\!+\!\theta$ for $s_t<-\theta$. In the asymmetric models, there exists only a single positive threshold $\theta>0$. Here, without added noise, the sensor output is zero for $s_t<\theta$ and $s_t\!-\!\theta$ for $s_t>=\theta$. We note that the analytical model described above belongs to the class of discrete symmetric models.
As one example of the detectors with memory, were the output depends not only on the momentary input and added noise, but also on earlier internal states of the detector itself, we choose the leaky integrate-and-fire neuron model \cite{lapique1907recherches,burkitt2006review} with
\begin{eqnarray}
\dot{x}&=&-\frac{1}{\tau_m}x+s_t\nonumber\\
\end{eqnarray}
where $x$ is the membrane potential, $\tau_m$ the mebrane time constant and $s_t$ the input signal.
If $x$ crosses the threshold $\theta$ from below, an output spike is generated and the membrane potential is set to the resting potential $x_{r}$, which is chosen to be zero for simplicity.

\onecolumn


\begin{figure}[htb]
\centering
\includegraphics[width=0.75\linewidth]{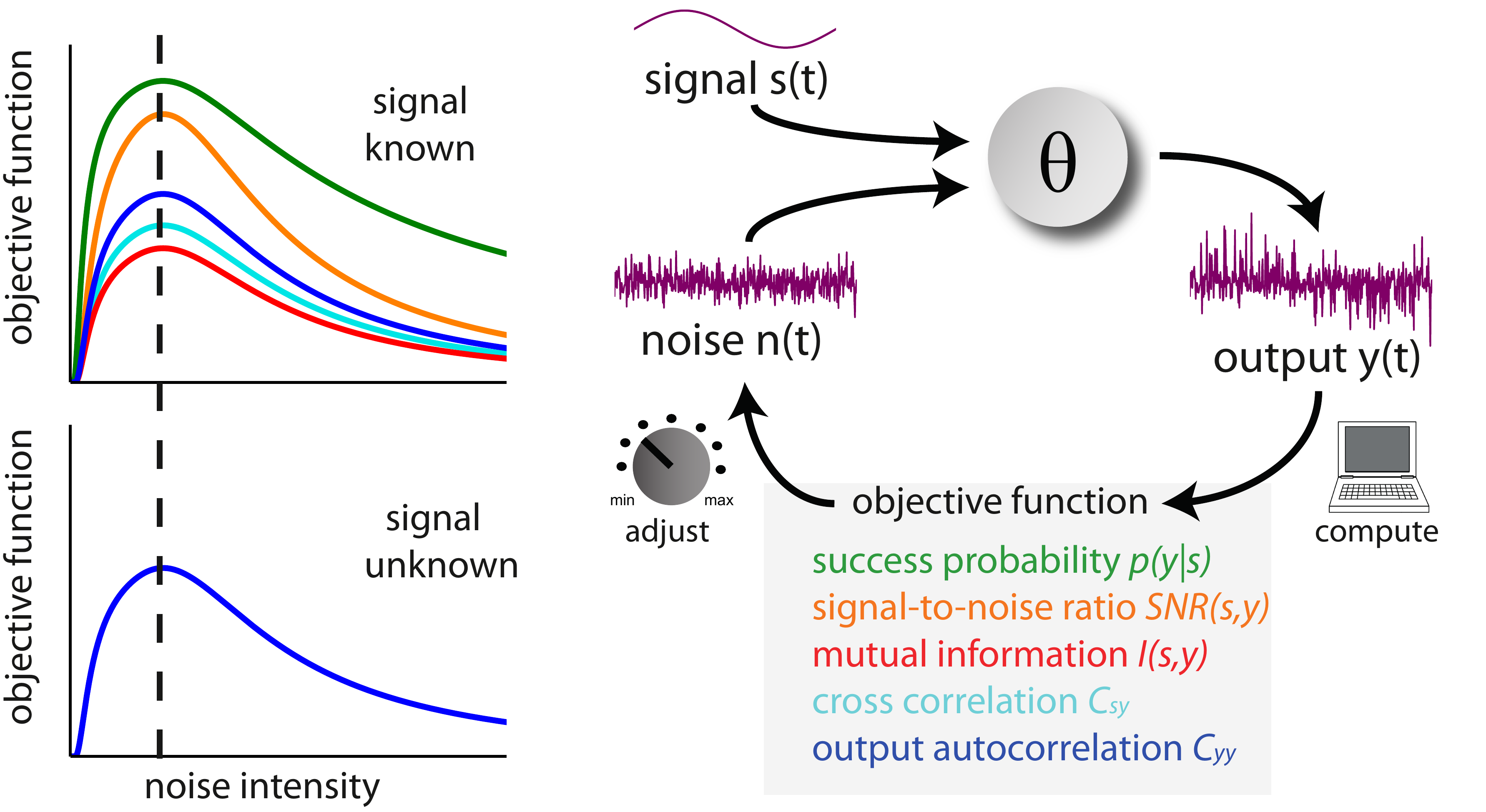}
\caption{\em \small Sketch of the adaptive stochastic resonance principle. In case, where the underlying signal is known, all objective functions can be calculated, whereas if the signal to be detected is unknown the only objective function that can be computed is the output autocorrelation.}
\label{fig:1}
\end{figure}

\begin{figure}[htb]
\centering
\includegraphics[width=1.0\linewidth]{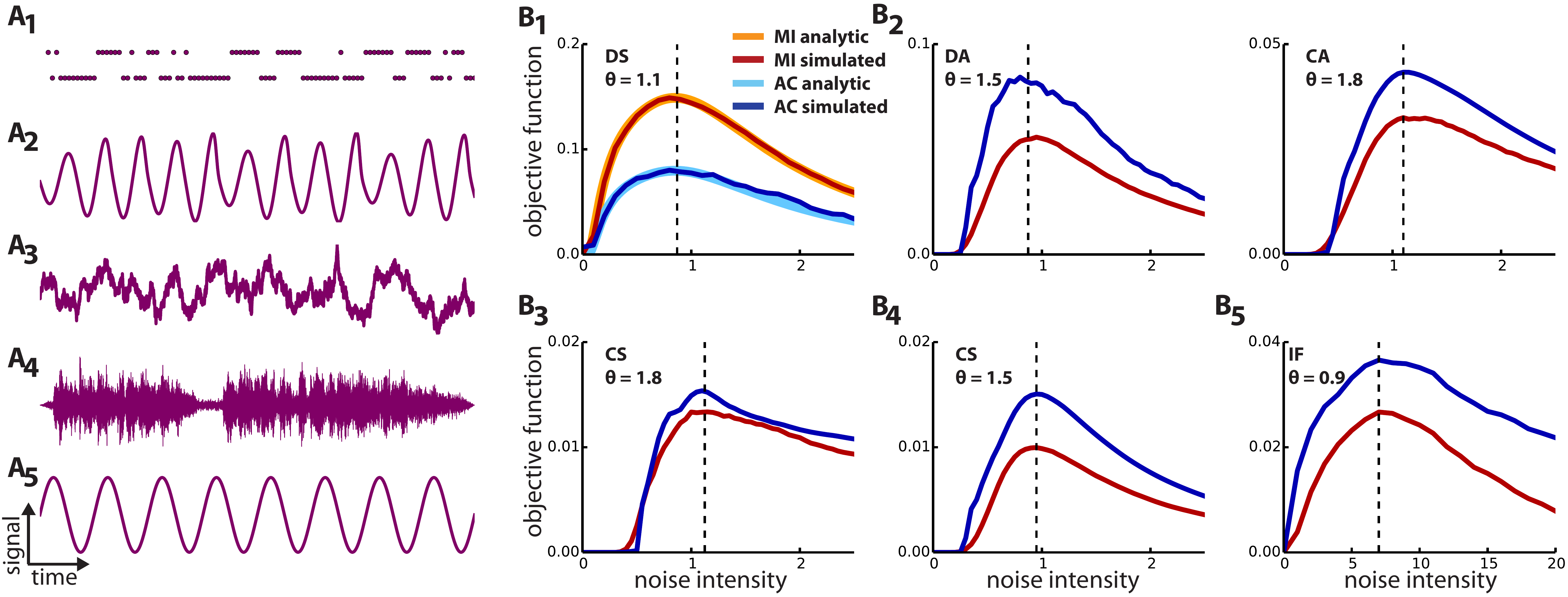}
\caption{\em \small A: Sample signals used for numerical simulations. As a discrete signal a correlated bipolar chain (A1) was chosen. As continouos aperiodic signals served, in order of complexity, one dimension of the 3-dimensional Roessler attractor (A2), an Ornstein-Uhlenbeck process (A3) and a wave file of recorded speech (A4). A simple sinosoid signal (A5) was generated as continuous periodic input. B: Sample results of numerical simulations, i.e. resonance curves of mutual information (red lines) and output autocorrelation (blue lines) for different models. The numbering corresponds to that of subfigure (A) and indicates the underlying input signals. The letters refer to the detector models with continuous/discrete output (C/D) and asymmetric/symmetric threshold (A/S). The case shown in (B1) corresponds to the analytical model. In addition to the numerical results, the analytical predictions of mutual information (orange line) and output autocorrelations (cyan line) are plotted. (B5) is a sample result of the integrate-and-fire neuron model.}
\label{fig:2}
\end{figure}

\begin{figure}[htb]
\centering
\includegraphics[width=0.8\linewidth]{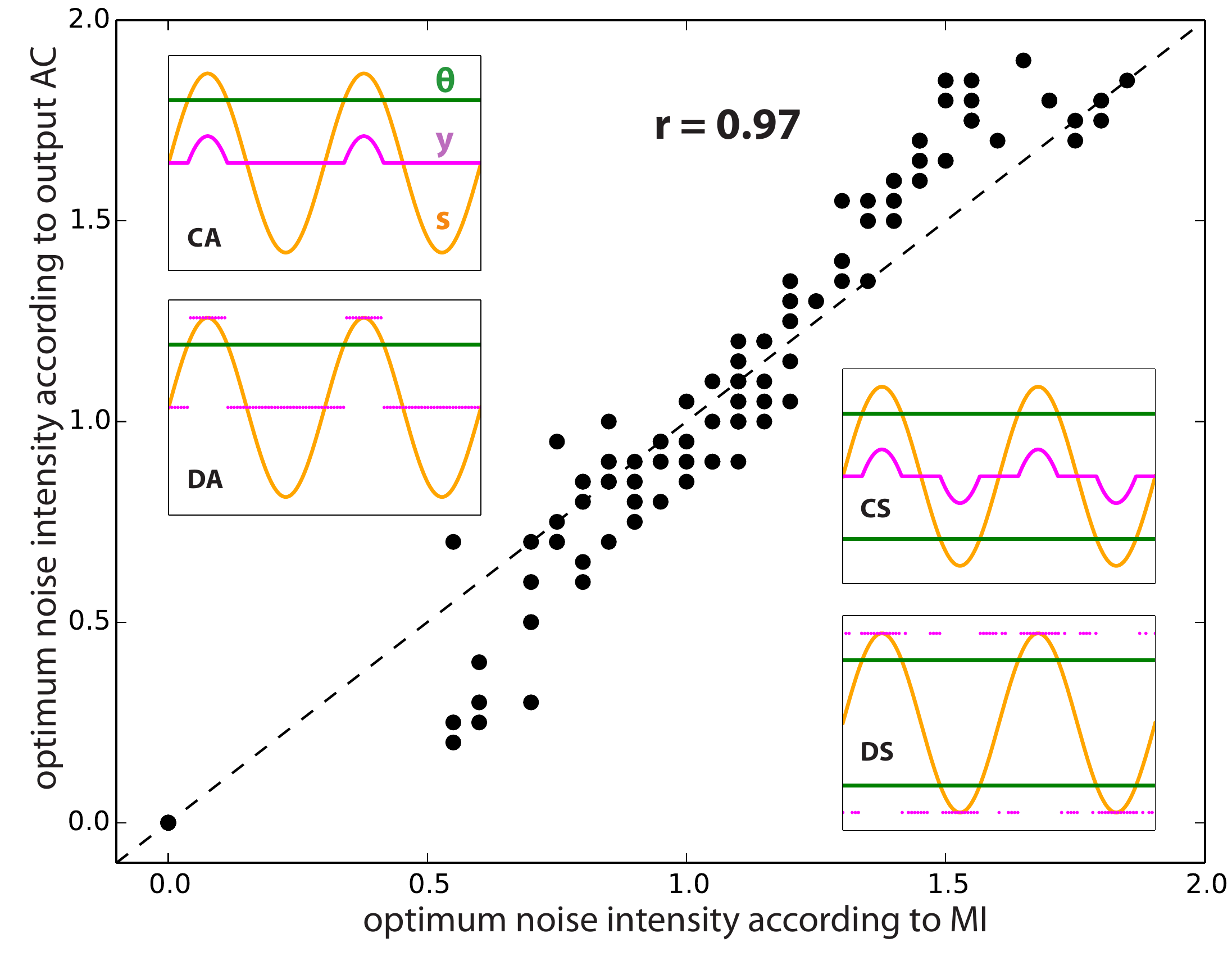}
\caption{\em \small Scatter plot that summarizes a large number of investigated models comprising different signal types, different detector types and different signal-to-threshold distances. The optimal noise intensities according to mutual information (MI) are plotted versus the corresponding intensities found by maximizing the output autocorrelation (AC). Remarkably both measures are highly correlated (correlation coefficient $r=0.97$).}
\label{fig:3}
\end{figure}

\twocolumn

\bibliographystyle{plain}
\bibliography{refs}

\end{document}